\documentclass[acmsmall]{acmart}
\AtBeginDocument{%
  }

\setcopyright{acmlicensed}
\copyrightyear{2025}
\acmYear{2025}
\acmDOI{XXXXXXX.XXXXXXX}
\acmConference[Conference acronym 'XX]{Make sure to enter the correct
  conference title from your rights confirmation email}{May 1,
  2025}{College Station, TX}
\acmISBN{978-1-4503-XXXX-X/2025/02}

\begin{document}

\title{Destructive Interference: Encoding Loss in the Overlap}

\author{Nikolas Aberle}
\email{nik.aberle@tamu.edu}
\orcid{1234-5678-9012}
\affiliation{%
  \institution{Texas A\&M University}
  \city{College Station}
  \state{Texas}
  \country{USA}
}

\renewcommand{\shortauthors}{Aberle}

\begin{abstract}
Destructive Interference is a data visualization installation that representing the deaths and injuries caused by mass shootings in 2024 in the United States. I parametrically designed and fabricated an interlocking ring sculpture for each month of 2024; where the overall height corresponds to the level of violence in that month. Taller forms mark the deadliest months, while shorter ones reflect fewer casualties. Each inner ring encodes the number of people killed or injured, and each outer ring encodes the number of shootings and the number of days without them. The interlocking cylinders are powered via a motor to rotate, and lit from within. As the cylinders rotate, they cast overlapping shadows that represent those killed or injured by mass shootings. The goal of this work is to visualize otherwise overwhelming and disparate statistics in a way that is both physically present and emotionally resonant. By inviting viewers to step into and engage with these shadows, the piece creates space for reflection, conversation, and confrontation with the scale of this ongoing crisis.
\end{abstract}


\begin{teaserfigure}
  \includegraphics[width=\textwidth]{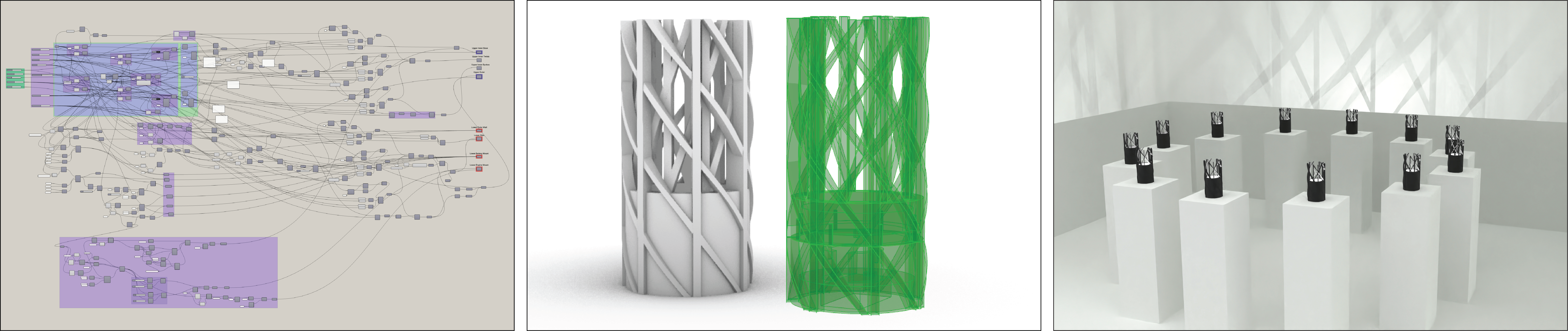}
  \caption{My parametric algorithm in Grasshopper used to create the interlocking cylinders (left), the surfaces in green and baked geometries in white in Rhino (center), and a potential render of the final installation (right).}
  \Description{My parametric algorithm in Grasshopper used to create the interlocking cylinders (left), the surfaces in green and baked geometries in white in Rhino (center), and a potential render of the final installation (right).}
  \label{fig:teaser}
\end{teaserfigure}

\received{20 February 2007}
\received[revised]{12 March 2009}
\received[accepted]{5 June 2009}

\maketitle

\section{Introduction And Related Works}
In a previous project I developed a parametrically designed Moire pattern generator in Grasshopper. I wanted to take that process and expand upon it, specifically in the direction of application to physical sculpture. Big data has been integrated into contemporary sculpture in a variety of ways and methods \cite{bourgault2023coilcam} and this process opens up a new avenue for me to work within, using data as its own medium.

I chose to create a data visualization sculpture that falls into the category of symbolic representation \cite{vande2010physical} according to the writings of Andrew Vande Moore \& Stephanie Patel whose paper on the categorization of physical visualization of information guided my design choices. Symbolic representation is defined by them “to be works characterized by a form-finding approach that employs a so-called arbitrary data-mapping approach”.

I chose this method as I wanted to visually disentangle the topic of mass shootings from a literal representation. I wanted people to engage with the work before being presented with the data linked to the creation so they were less likely to disengage as soon as they realized what the subject matter was.

Similar to how I created my previous Moire disks, I designed this project in Grasshopper to be baked in Rhino after the final formulation of each month's data. One challenge of this was going to be making the design as minimal as possible as to not draw attention away from the aforementioned meaning. My goal was to create a print in place upper piece that didn’t require any after assembly, but still allowed for rotation; for this I referenced existing methods of 3D printing interlocking parts \cite{song2015printing}.

According to the Gun Violence Archive, an online archive of gun violence incidents collected from over 7,500 law enforcement, media, government and commercial sources daily \cite{gunviolencearchive2024}, there were 586 mass shootings in 2024; resulting in a total death count of 711 individuals.

\section{Methodology}

As a first step, I collected and organized mass shooting data for the 2024 calendar year from the Gun Violence Archive \cite{gunviolencearchive2024}, accessed originally through Wikipedia \cite{wikipedia2024massshootings}. This included the number of shootings, total injuries and deaths, and the number of days without a mass shooting for each month.

\begin{figure}[H]
  \centering
  \includegraphics[width=\textwidth]{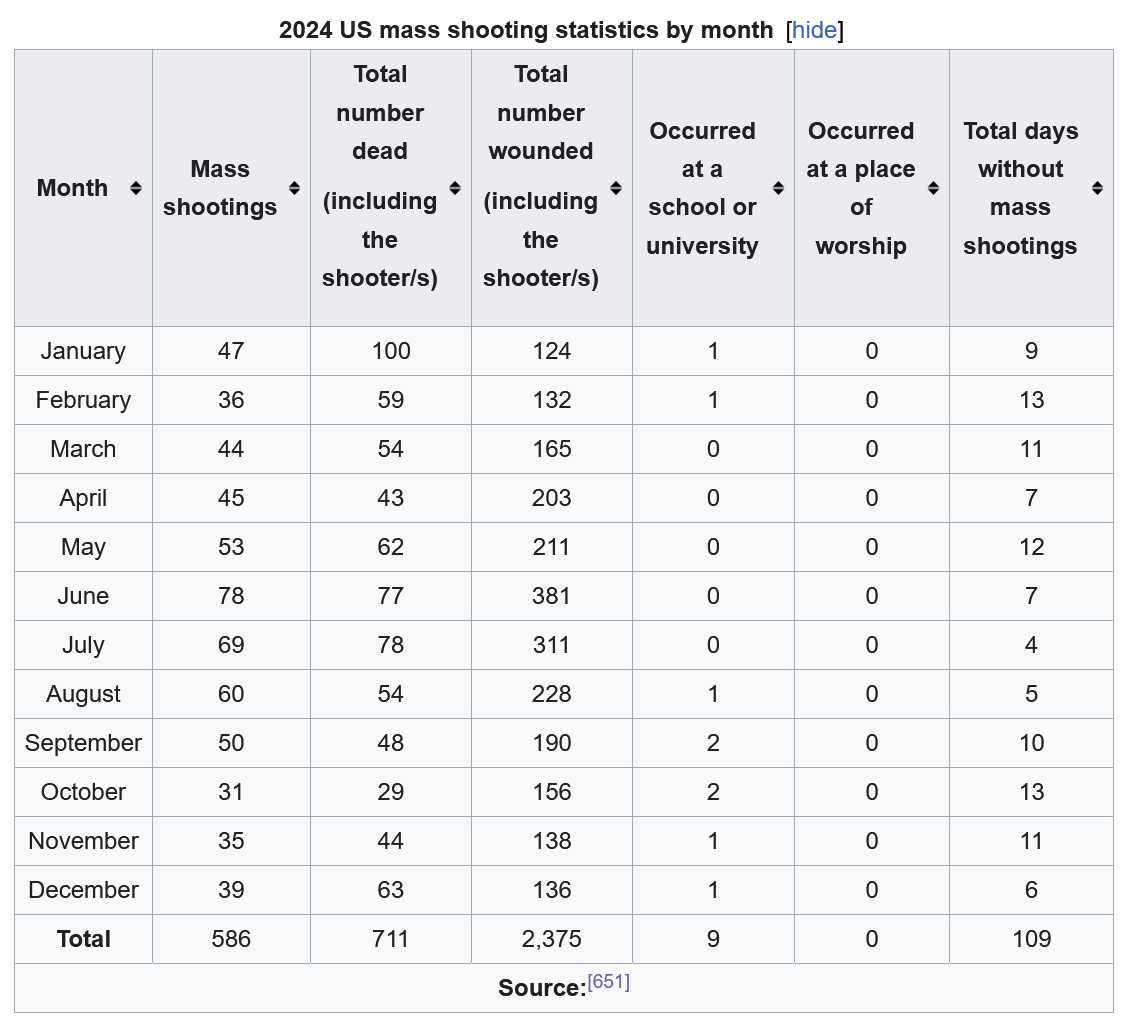}
  \caption{2024 US mass shooting statistics by month per the Gun Violence Archive.}
  \Description{2024 US mass shooting statistics by month per the Gun Violence Archive.}
\end{figure}

I used this data as the main inputs to a parametric model in Grasshopper. Each of the twelve sculptures was generated from the same base algorithmic design: a set of two cylinders interlocked at the base, but still allowing for rotation. I wanted to encode the data from the 2024 shootings into specific parameters that would carry throughout each sculpture.

The heights of the sculptures are reflective of the total number of deaths in each month, normalized between 8” for the highest death count and 3” for the lowest death count. Both inner and outer rings have a number of straight vertical pillars and pillars that rotate around the cylindrical shape based on a rotational value.

The number of vertical pillars on each inner ring corresponds to the ratio of deaths in that month to total deaths in 2024, and the rotational value of the twisted spokes is total wounded for that month compared to total wounded and normalized to a rotational value of 180.

\begin{figure}[H]
  \centering
  \includegraphics[width=\textwidth]{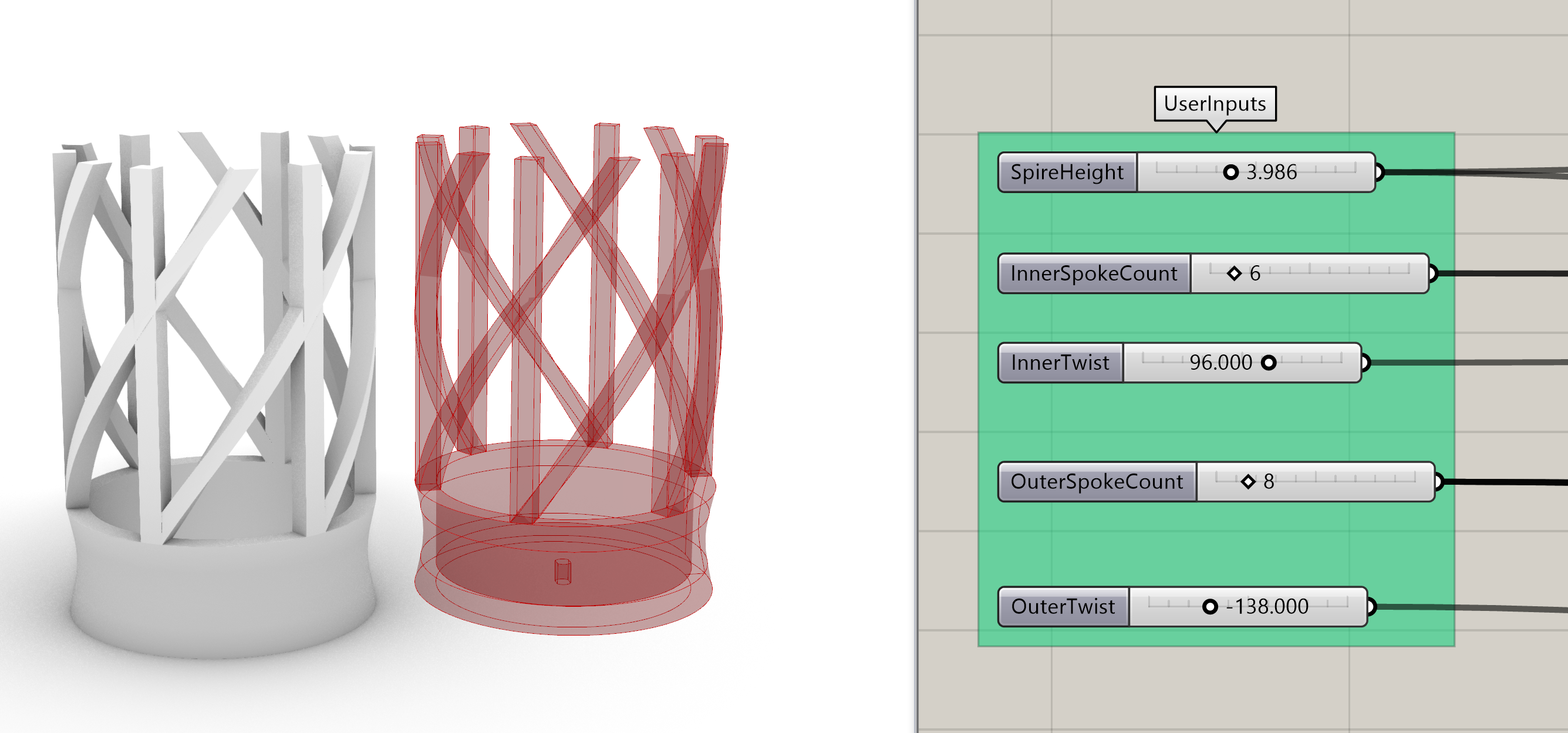}
  \caption{Real data from April 2024. InnerSpokeCount set to 6 represents 6\% of the deaths from 2024 being in April. The InnerTwist set to 96 represents the number injured in April in relation to the total number injured and that number normalized to 0 - 180 degrees.}
  \Description{Real data from April 2024. InnerSpokeCount set to 6 represents 6\% of the deaths from 2024 being in April. The InnerTwist set to 96 represents the number injured in April in relation to the total number injured and that number normalized to 0 - 180 degrees.}
\end{figure}

The number of vertical pillars on each outer ring corresponds to the ratio of shootings in that month to total shootings in 2024, and the rotational value of the twisted spokes is the number of days without a shooting in that month and normalized to a rotational value of -180.

\begin{figure}[H]
  \centering
  \includegraphics[width=\textwidth]{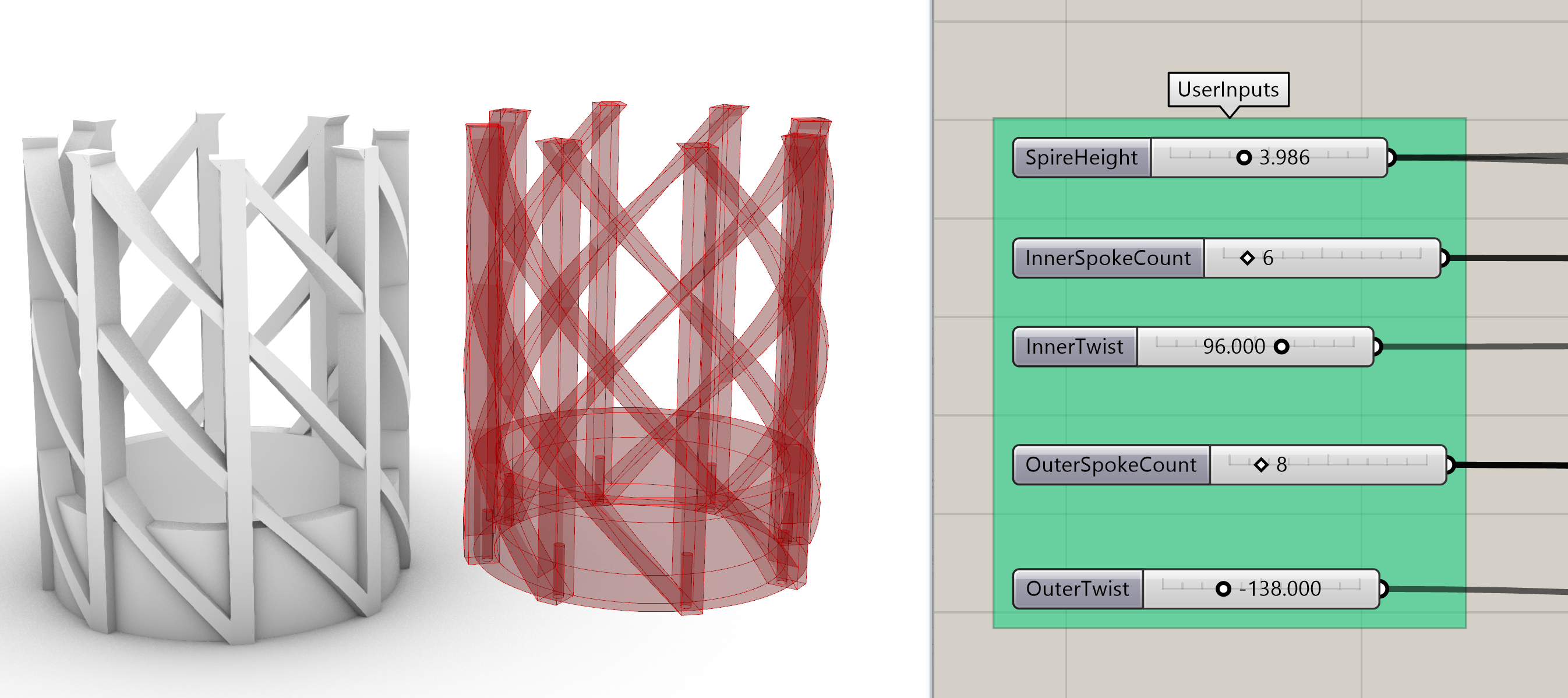}
  \caption{Real data from April 2024. OuterSpokeCount set to 8 represents 8\% of the mass shootings from 2024 being in April. The OuterTwist set to -138 represents the number of days without shootings in April in relation to the total number of days in April and that ratio normalized to -180 to 0 degrees.}
  \Description{Real data from April 2024. OuterSpokeCount set to 8 represents 8\% of the mass shootings from 2024 being in April. The OuterTwist set to -138 represents the number of days without shootings in April in relation to the total number of days in April and that ratio normalized to -180 to 0 degrees.}
\end{figure}

The inner ring rotation being 0 - 180 and the outer ring rotation being -180 - 0 was an intentional design choice to amplify the shapes of the shadows.

\begin{figure}[H]
  \centering
  \includegraphics[width=\textwidth]{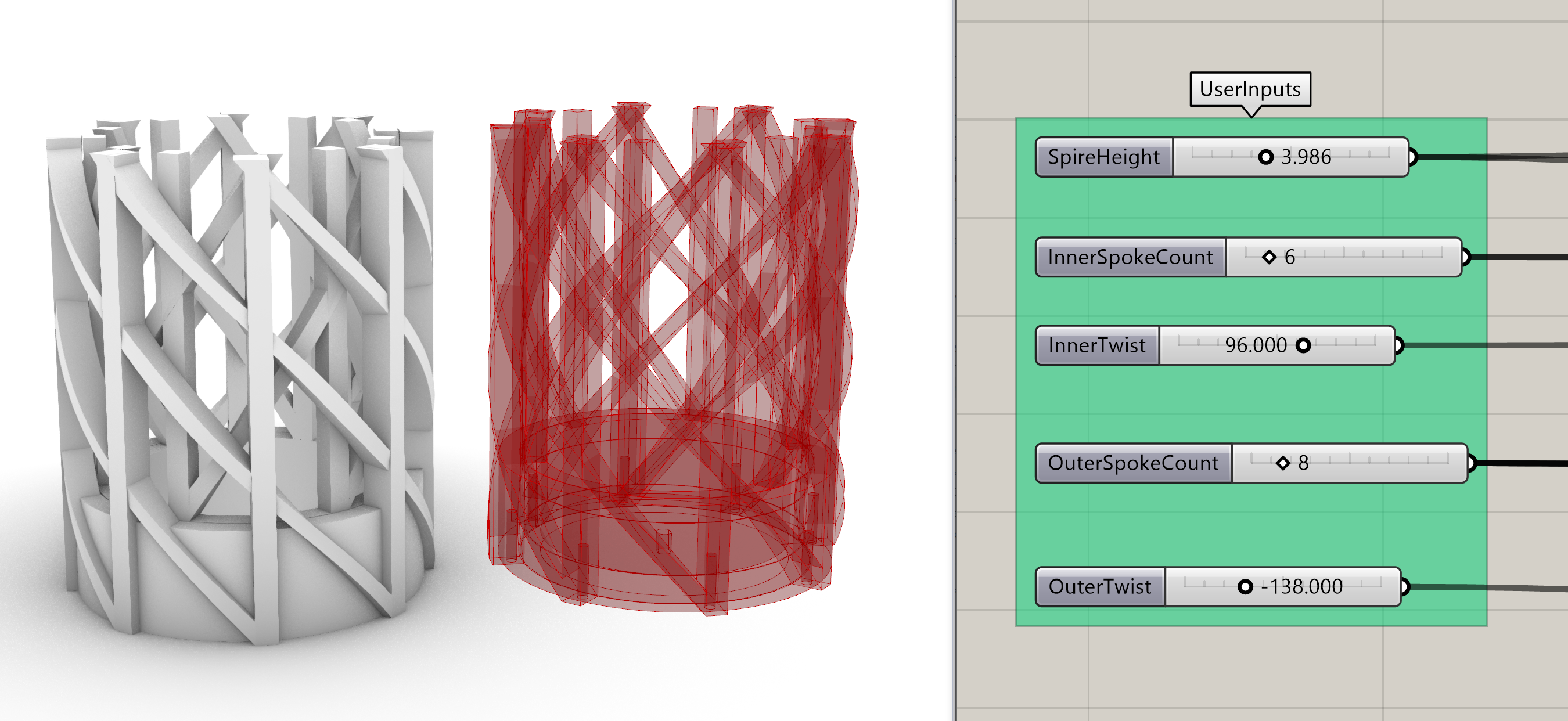}
  \caption{The final representational sculpture for April 2024. To calculate spire height I assigned the month with the most deaths (January) to 8" and the month with the fewest deaths (October) to 3", and mapped the rest of the months to their corresponding values within that range. April had the second lowest deaths of 2024.}
  \Description{The final representational sculpture for April 2024. To calculate spire height I assigned the month with the most deaths (January) to 8" and the month with the fewest deaths (October) to 3", and mapped the rest of the months to their corresponding values within that range. April had the second lowest deaths of 2024.}
\end{figure}

I deliberately avoided using overt graphing or labeling, as the intent was to present the data in a form that emphasized presence over explanation, in line with the symbolic representation definition. After the final geometry for each month was generated it was baked in Rhino and 3D printed.

The last step was to create a base design that allowed me to mount a battery pack and 6v 5rpm turntable motor in it to mount each upper part to. The outer ring mounts to the base with 8 alignment pins and ensures the inner ring can rotate freely when the motor is powered. A 3” LED puck is then placed in the center and the sculpture is ready to be installed along the other 11.

\begin{figure}[H]
  \centering
  \includegraphics[width=\textwidth]{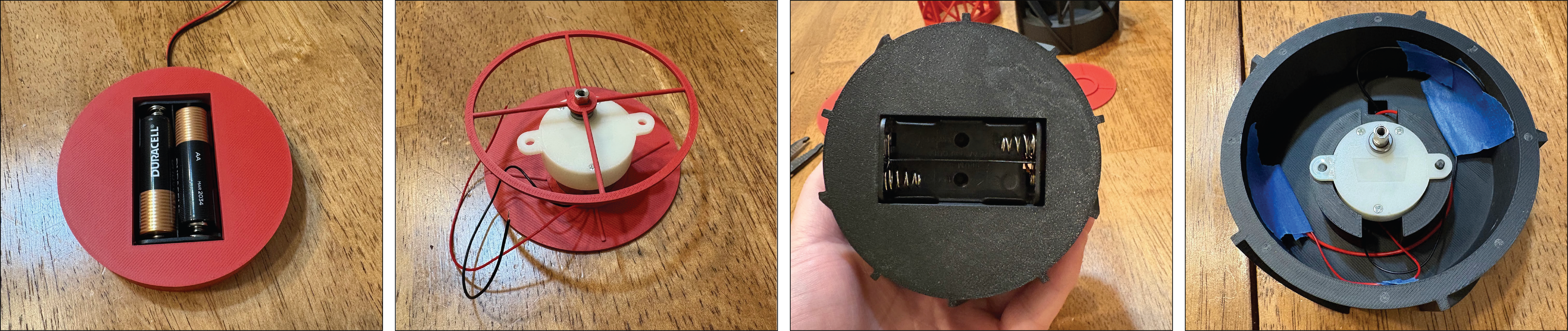}
  \caption{First pass at the battery mount (left), first pass at motor centering (second from left), final battery mount in base (second from right), final motor mount in base (right).}
  \Description{First pass at the battery mount (left), first pass at motor centering (second from left), final battery mount in base (second from right), final motor mount in base (right).}
\end{figure}

When all 12 are placed in proximity to one another, walking between them and engaging with the shadows places you amidst all of the loss due to mass shootings in 2024.

\section{Result And Future Work}

The twelve individual sculptures for each month have been fully designed, fabricated, and tested individually. While the full installation has not yet been fully installed in a final venue, I’ve tested six of the units running simultaneously, and the results align closely with my vision and goals. The slow, overlapping motion and cast shadows work as intended, creating a layered and immersive experience that invites engagement.

\begin{figure}[H]
  \centering
  \includegraphics[width=\textwidth]{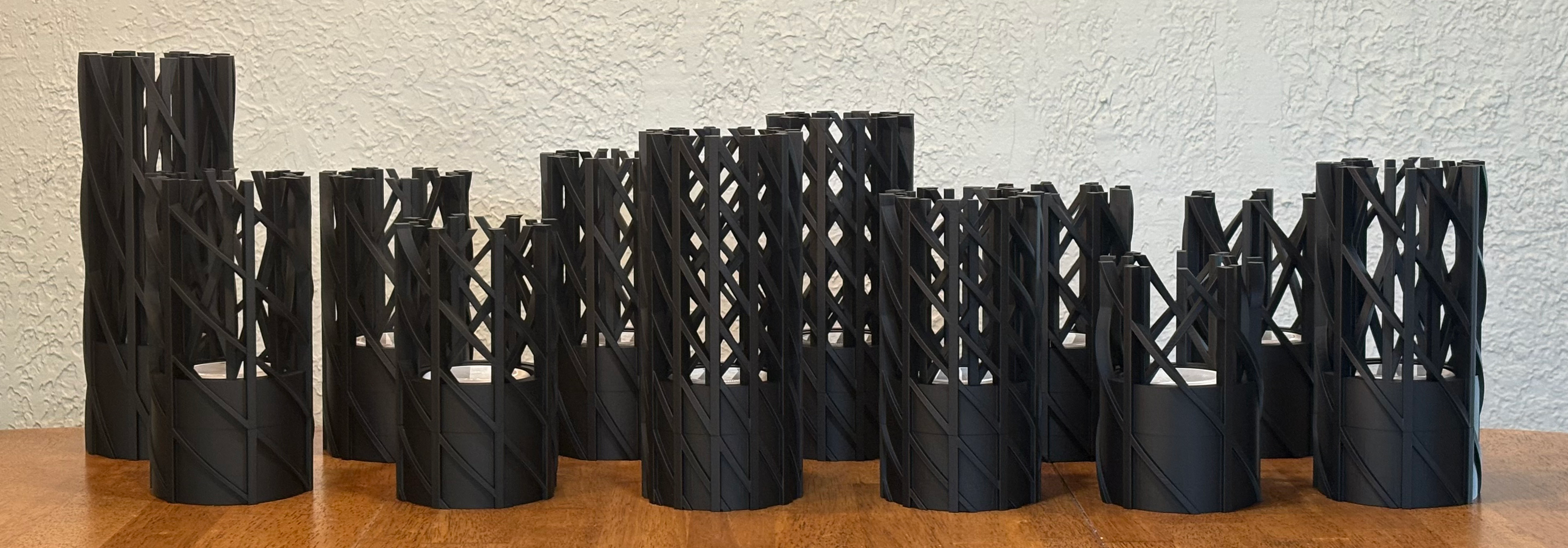}
  \caption{All twelve months on display in a zig zag pattern, with January in the back left, and December in the front right.}
  \Description{All twelve months on display in a zig zag pattern, with January in the back left, and December in the front right.}
\end{figure}

The 3D interlocking rings function mostly as intended, though there is the occasional snag or slowdown of the motor. More investigation on the cause of this will be needed in the future. Additionally, I may need to look at using four AA batteries instead of two as the motor tends to stall even under minimal load. I plan to also look into the possibility of using Lithium-ion polymer batteries in the future, as they last longer and would be easier to recharge for prolonged displays of the work. Once a full installation space is available, I plan to finalize the layout to best encourage viewer movement and interaction.

For future projects I’d like to explore responsive elements tied to real-time or streamed data, as well as other methods of converting large bodies of data around socially charged topics into symbolic representations for easier engagement. This project has laid the groundwork for future installations that invite sustained, embodied reflection on topics that are often too vast or diffuse to confront directly.

\section{Conclusion}

This project was something I’ve wanted to do for a long time, taking concrete data and using it to inform the message and form of an installation. In this case it has the added benefit of being something emotionally charged and allows for viewers to hold space for both reflection and discomfort. The two most memorable moments were when my print in place interlocking rings functioned as intended and I just had to dial in the tolerances. Second was when the first full sculpture (January) was assembled, powered, and ran as intended; casting an amazing cascade of shadows on my studio walls.

Through this process I learned how to use data to inform the parameters of a sculpture in a meaningful way, as well as how viewers can interact with it once created. My hope is that this work will encourage deeper reflection on statistics that are often overlooked or diminished; and in this specific case to bring more awareness to the toll mass shootings take on the American populace.

\begin{acks}
This is part of the Texas A\&M course - VIZA 626. Assignment 3, during the Spring 2025 Semester.
\end{acks}

\bibliographystyle{ACM-Reference-Format}
\bibliography{sample-base}
\end{document}